# Structural, electronic and magnetic properties of the Manganese telluride layers AMnTe$_2$ (A=K, Rb, Cs) from first-principles calculations

A. Benmakhlouf[1,2], Y. Bourourou[3], A. Bouhemadou[4], A. Bentabet[2], F. Khemloul[5], S. Maabed[5], M. Bouchenafa[5], I. Galanakis[6*]

[1]Département de Sciences Technique, Faculté de Technologie, université de Laghouat, Algérie

[2]Laboratoire de Recherche : Caractérisation et Valorisation des Ressources Naturelles, Université de Bordj Bou Arreridj, 34000, Algérie

[3] Laboratoire de Modélisation et Simulation en Sciences des Matériaux (LMSSM) Université Djilali Liabes de Sidi Bel Abbes 22000 / Algérie

[4] Laboratory for Developing New Materials and their Characterization, University of Setif 1, Setif 19000, Algeria

[5]Faculty of sciences, Sciences de la matière département, Laghouat University, 03000, Algeria

[6] Department of Materials Science, School of Natural Sciences, University of Patras, Patras 26504, Greece

*Corresponding author: galanakis@upatras.gr

*Abstract*

Using first-principles electronic structure calculations based on density functional theory (DFT), we investigate the structural, electronic and magnetic properties of the layered ternary manganese tellurides: AMnTe$_2$ (A = K, Rb, and Cs). Calculations are accomplished within the full-potential linearized augmented plane wave (FP-LAPW) using the generalized gradient approximation GGA formalism for the exchange correlation term. We have treated all ferromagnetic, antiferromagnetic and non-magnetic phases and found that the ferromagnetic is the ground-state for all studied compounds. Moreover, all three compounds under study are half-metals with a total spin magnetic per formula unit of $4\,\mu_B$ which is mainly localized at the Mn atoms. We express for these compounds a new version of the Slater-Pauling rule and discuss in detail the origin of the minority-spin gap. Finally, we have also calculated some other relevant quantities such as the bulk modulus B, the pressure derivative B', the virtual semiconducting gap E$_g$, and the half-metallic gap E$_{HM}$.

## 1. Introduction

The appearance of Half-metallic materials, in early 1980 [1, 2], has paved the way for a new technology in field of electronics. Since their discovery, these compounds are becoming more and more interesting as an entity allowing the introduction of new concepts in the storage and processing of information in the context of spintronic [3-5]. A half metal is a solid with a strange electronic structure. The fermi level intersects only one of the two spin bands (metallic behavior) while for the other spin direction there is a gap just like in semiconductors or insulators. In the ideal case, this mixed property makes it possible to reach a spin polarization of 100% at the Fermi level, and thus has the advantage of maximizing the efficiency of spintronic devices [6]. Nowadays, growth techniques allow the synthesis of novel materials and a large number of these compounds are identified to be Half–metals. Among them we can quote: The Heusler and full Heusler materials [7, 8], double perovskites [9], zinc-blende compounds [10], thiospinel systems [11], transition metal pnictides [12] and some metallic oxides [13, 14]. Two recent theoretical studies have predicted the half-metallicity in ternary tetragonal structures having the chemical formula $AMnQ_2$, were A is an alkali metal and Q is a chalcogenide atom [15, 16].

Among A-Mn-Q compounds, the $AMnTe_2$ (A= K, Rb, Cs) systems have not been sufficiently explored which is, perhaps, related to a significant extent to technical difficulties in growing them [17]. In the present article we systematically investigate the structural, electronic and magnetic properties of the ternary layered compounds $KMnTe_2$, $RbMnTe_2$ and $CsMnTe_2$, synthetized for the first time by Kim *et al* [17]. Since 1999, this work remains, to the best of our knowledge, the only published article devoted to these compounds. Authors in reference [17] reported the crystal structure and parameters of these three materials and have also obtained the temperature-dependent molar magnetic susceptibility and measured their effective moment.

This study is dedicated to investigate theoretically the $AMnTe_2$ (A=K, Rb, Cs) compounds and therefore provide interesting information on their physical properties as a reference for future studies. The article is organized as follows: In Section 2 we describe the computational details used in this work. The obtained results are reported and discussed in Section 3. Finally, conclusions and remarks are provided in Section 4.

## 2. Calculation details

To perform the calculations we applied the first principles full-potential linearized augmented plane wave (FPLAPW) method which is based on the density functional theory (DFT) [18, 19] as implemented in the Vienna package WIEN2K [20]. For the exchange correlation potential, we used the generalized gradient approximation (GGA) in the Perdew-Burke-Ernzerhof parameterization which is well-known to produce accurate equilibrium lattice parameters for half-metallic compounds [21].

Within this method the space is divided into non-overlapping muffin-tin (MT) spheres which are separated by an interstitial region. The Kohn–Sham equations are solved self-consistently and the wave functions are expanded in terms of spherical harmonic functions inside the muffin tin spheres and in terms of Fourier series in the interstitial region. For the expansion of the wave functions in the interstitial region, the value of the cut-off parameters was chosen as $R_{MT} \times K_{max} = 8.0$ where $R_{MT}$ is the muffin-tin sphere radii and $K_{max}$ is the largest reciprocal lattice vector. For the wave function expansion inside the muffin-tin spheres, the maximum value of the orbital momentum (lmax) was assumed equal to 10. For the $KMnTe_2$ compound, the muffin-tin sphere radii $R_{MT}$ was chosen as to 2.4 au for all three K, Mn and Te atoms, while for the $RbMnTe_2$ and $CsMnTe_2$ compounds we assumed a value of 2.0 a.u for all Rb, Cs, Mn and Te atoms. The separation energy of valence and core states (cut-off energy) was chosen as -6.0 Ry. To perform the integrations in the reciprocal space we considered 2000 k-points in the Brillouin zone (BZ). We assumed that the self-consistent calculations are converged when the integrated charge difference per formula unit between the input charge density and the output charge density is less than 0.0001 in arbitrary units.

## 3. Results and discussion
### 3.1. Total energy and crystal structure

The X-ray powder data of the $AMnTe_2$ (A=K, Rb, Cs) compounds have been reported in literature [17]. The title materials have a dark-red color and are isostructural with $AgTiTe_2$-type crystal structure, space group I-4m2 (No. 119) of the tetragonal system, with 2 formula units (i.e. 2 $AMnTe_2$) per unit cell. The unit-cell of these materials is presented in Fig. 1. As illustrated, these tetragonal structures can be regarded as a layered material, where each [$MnTe_2$] layer is separated by alkali atoms that are at the center of slightly compressed square prisms formed by 8 Te atoms. Interlayer bonding is formed between A and Te atoms while the one between Mn and Te atoms is an intralayer bonding.

The structural properties of the $AMnTe_2$ systems in non-magnetic (NM), ferromagnetic (FM) and antiferromagnetic (AFM) configurations are determined by calculating the total energy of several different volumes using GGA. The calculated total energies versus unit cell volume were fitted to the Murnaghan equation of state [22]. The obtained results are shown in Fig. 2 and they indicate that for all studied materials the ferromagnetic phase is the energetically favorable ground state. The calculated crystal data and experimental details of the equilibrium lattice constants and atomic positions are listed in Tables 1 and 2, respectively. Our calculated lattice constants for the $KMnTe_2$ and $RbMnTe_2$ compounds are in good agreement with the experimentally determined lattice parameters reported by J. Kim et al. [17]. Contrary to these compounds, in the case of $CsMnTe_2$ in the FM phase, the calculated lattice parameters *a* and *c* and the *c/a* ratio deviate 4.1%, 16.8% and 36%, respectively, from the measured ones. This discrepancy for $CsMnTe_2$, may be attributed to the importance played by the Van der Waals interactions between successive layers which are not taken into account in the electronic structure method which we have used.

The obtained bulk modulus B and its pressure derivative B' from the E versus V curve fitted to the Murnaghan equations of state are listed in Table 1. To the best of our knowledge, these are the first predicted values of the B and B' for the $AMnTe_2$ compounds. These results show a similarity with other theoretical results for other isostructural structures [15]. With these relatively low B values, all studied materials can be characterized as very soft materials.

In the following we are interested only in the study of the ferromagnetic (FM) phase of the studied materials.

### 3.2. Electronic properties

In Figs. 3-5 we present the calculated spin-polarized band structure along several high-symmetry directions of the first Brillouin zone in the ferromagnetic phase of the studied materials at the theoretical equilibrium lattice structure. As shown, for all compounds, the fermi level $E_F$ crosses the majority-spin (spin-up) bands showing a metallic nature, and it falls in a band gap of 1.29 eV, 1.37 eV and 1.25 eV for $KMnTe_2$, $RbMnTe_2$, and $CsMnTe_2$, respectively, for the minority-spin (spin-down) direction revealing their semiconducting nature. According to this, the conduction electrons at the Fermi level are thus fully spin polarized and the $AMnTe_2$ (A= K, Rb, Cs) compounds are half-metallic ferromagnets. The minority-spin band gaps are found to be indirect between the Γ point of the highest valence band and the X point of the lowest conduction band. The majority-spin bands crossing $E_F$ are rather dispersive, and the densities of states at the Fermi level are 1.68, 1.71 and 1.54 states per eV per spin per unit-cell for $KMnTe_2$, $RbMnTe_2$, and $CsMnTe_2$, respectively. The planarity of some of the majority-spin bands means that Mn-*d* electrons are playing a dominant role [1]. The spin-flip gap (the half-metallic gap $E_{HM}$) which is defined as the minimum energy required to flip a majority-spin electron from the valance band maximum to the minority-spin Fermi level [23], is 0.55 eV, 0.65 eV and 0.57 eV for $KMnTe_2$, $RbMnTe_2$, and $CsMnTe_2$, respectively.

We have also calculated the spin-projected total and partial density of states (DOS), to further explain the electronic band structure. The corresponding partial densities of states (DOS) are illustrated in Figs. 6-8. Accordingly, the calculated band structures show that the occupations of the K-s, Rb-s and Cs-s states are very small. The absence of K, Cs, and Rb partial DOS in the valence-band region (their p occupied states are lower than the continuum shown in the DOS figures) shows that the role of the A alkali metals in this compound is restricted to donating one electron to the $AMnTe_2$ host band structure and keeping the $AMnTe_2$ sandwiches apart [24]. The lowest energy bands at -10.12 eV and from -10.91 eV to -10.07 eV and from -10.14 eV to -8.29 eV for $KMnTe_2$, $RbMnTe_2$ and $CsMnTe_2$, respectively, are derived from an admixture of A-p and Te-s states. Near the fermi level similar profiles of the densities of states are noted in all compounds, with a difference in the value of the gap in the spin-down states. Examination of Figs. 6-8 reveal that the densities of states near the fermi level ranging from about -4.0 eV to 0.90 eV for the spin-up states and from about -3.50 eV to 2.50 eV for the spin-down states are formed mainly by the transition metal Mn-d states and Te-p orbitals for all compounds.

The energy dependence of the electron spin polarization $P_e$ (E) is given by

$$P_e(E) = \frac{n_\uparrow(E) - n_\downarrow(E)}{n_\uparrow(E) + n_\downarrow(E)}$$

where $n_\uparrow(E)$ and $n_\downarrow(E)$ are the spin dependent density of states. The electrons at the Fermi energy are fully spin polarized ($P_e = 100\%$) for all three compounds since $n_\downarrow(E_F) = 0$ as shown in Figs. 6-8.

### 3.3. Magnetic properties and Slater-Pauling rule

The calculated values of the total spin magnetic moment per formula unit and the partial magnetic moments are listed in Table 3. The total magnetic moment per formula unit (f.u.) in $\mu_B$ is just the difference between the number of spin-up (↑) occupied states and the spin-down (↓) occupied states. Our calculations give for all compounds study a total spin magnetic moment per f.u. of 4 $\mu_B$. The integer value is expected since there is an integer number of minority-spin occupied bands due to the gap. Experimentally the total spin magnetic moments per f.u. measured were found to be 4.31 $\mu_B$, 4.71 $\mu_B$ and 4.86 $\mu_B$ for KMnTe$_2$, RbMnTe$_2$, and CsMnTe$_2$, respectively [17]. The discrepancy between the calculated values and the experimental ones may be related to the presence of impurities and defects or to not 100% crystallinity in the experimental samples. No definitive conclusions can be drawn regarding this discrepancy and further experimental investigations are needed.

Most of the total spin magnetic moment is concentrated at the Mn atoms. The situation is similar to the well-known and widely studied NiMnSb semi-Heusler compound [17]. The spin-magnetic moment is localized to the Mn atoms due to the exclusion of the minority-spin states below the Fermi level as shown in the Mn DOS presented in Figs. 6-9 for all three compounds. Almost all majority Mn d-states are occupied while almost all minority Mn d-states are unoccupied leading to a large value of the exchange splitting of the d-states (energy difference between the mass-center of the minority and the mass-center of the majority spin bands). Large values of the exchange splitting mean robust ferromagnetism which is stabilized due to ferromagnetic Ruderman-Kittel-Kasuya-Yosida interactions as discussed in detail in Refs. [6, 15, 25].

### 3.4. Slater-Pauling rule and origin of the minority-spin gap

As it is well known in half-metallic compounds the total spin magnetic moment per f.u. follows the so-called Slater-Pauling behavior showing a linear variation with the total number $Z_t$ of valence electrons in the f.u. [6, 15, 25]. In the compounds under study the form of this rule is $\mu_{tot} = Z_t - 22$ as we will discuss in detail in the following. First, since the total spin magnetic moment per f.u. is 4 $\mu_B$, this means that there are in total 26 valence electrons per f.u. This is easily confirmed if we just count the number of valence electrons per atom. The alkali atoms K, Rb and Cs usually are referred to as monocations since they have one 4s, 5s or 6s valence electron, respectively. In the compounds under study we have also to consider as valence electrons also the six electrons accommodated in the fully occupied 4p, 5p or 6p orbitals, respectively, since as we will discuss in the next paragraph they are within the energy window considered for the valence electrons. Thus in total alkali atoms have 7 valence electrons. The Mn atom has also seven valence electrons (2 electrons occupy the 4s states and 5 electrons the 3d states). Each Te atom has six valence electrons, two occupying the 5s and four the 5p states. Thus in total we have 7+7+2x6=26 valence electrons per formula unit.

To fully elucidate all the aspects of the Slater-Pauling rule mentioned above we still have to discuss the origin of the number 22 which means that in the minority spin-band structure we have exactly 11 occupied states. We will base our discussion on the band structure of KMnTe$_2$ shown in Fig. 3 and we will expand it for the other two compounds using their band structures shown in Figs. 4 & 5. We have also performed a detailed fat band scheme analysis of the bands (not shown here), which reveals the orbital character of each band, to confirm our conclusions presented below. Of special importance is the character of the bands at the $\Gamma$ point which corresponds to the orbital character of the states in the real space. We should note that irrespectively of the orbitals which the valence electrons occupy in the free atoms, in the solids they occupy each time the lowest available electronic bands which may be of different orbital character as shown in Refs. [6, 15, 25]. Also, although the lattice is tetragonal, since the atoms sit at sites of tetrahedral symmetry, the symmetry arguments presented in Refs. [6, 15, 25] for the full-Heusler compounds are still valid taking into account that the z-axis is no longer equivalent to the x- and y-axis and thus for example the triple degenerate *p*-states will break into two double degenerate at the $\Gamma$ point ($p_x$ and $p_y$) states and a single $p_z$ state.

The energy window used in Fig. 3 contains all valence bands. At about -14 eV there is a completely flat band which shows no dispersion. This band stems from the K p-states and accommodates three electrons per spin and thus in reality it a triple-degenerate band. When we substitute Rb for K these specific bands move higher in energy and become dispersive being at the same energy position with the s-bands of Te described below. In the case of Cs they move even higher in energy than the s-states of Te being at -8 eV and become again pretty flat since the do not hybridize with other states. At about -10 eV in Fig. 3 there is a double degenerate band which stems from the s-states of Te atoms (there are two bands per spin because there are two Te atoms per formula unit). The energy position of these bands is almost the same for all three compounds.

The minority spin DOS below the Fermi level for all three compounds is dominated by the 5p-states of Te atoms with a very small admixture of the Mn 3d-states (if one analyzes further the DOS these d-states of Mn are of $t_{2g}$ and not $e_g$ character). This is expected since the p-states transform following the same irreducible representations with the $t_{2g}$ d-orbitals. Examining in detail the minority-spin band structure of KMnTe$_2$ in Fig. 3, we can see that there is one band at -4 eV at the Γ point followed by a double degenerate band. This set of bands is followed by a similar set of bands up to the Fermi level. These six bands form a continuum in DOS shown in Fig. 6 and are actually the p-bands of the two Te atoms (3 p-orbitals per spin and per atom) which have also a small admixture of the Mn $t_{2g}$ states. In cubic lattice they are triple degenerate at the Γ point but here they split to a single and a double degenerate one due to the tetragonal lattice as mentioned above.

Thus in total we have in the minority spin-band structure 11 occupied bands stemming from the 3 K, Rb or Cs p-, the 2 Te s-, and the 6 Te p-orbitals (for Te we take into account that there are two atoms per f.u.). Just above the gap in the minority-spin band structure we find the d-bands of the Mn atom. Actually, the minority-spin gap opens due to the hybridization of the Te p-states with the Mn $t_{2g}$ states which obey the same symmetry operations and which form bonding states below the Fermi level of mainly p-character and antibonding states above the Fermi level of mainly $t_{2g}$-character. The $e_g$ Mn d-states cannot hybridize with the Te p-states but they lie at almost the same energy window with the Mn $t_{2g}$ states above the Fermi level as shown by the partial DOS presented in Figs. 6-9.

### 3.5. Conclusion

In summary, using accurate density-functional electronic band-structure calculations we have made a systematic calculation of the structural, electronic and magnetic properties of the AMnTe$_2$ (A=K, Rb, Cs) layered compounds. We found that the ground state for all of them is the half-metallic ferromagnetic one. Therefore, if they would be grown in the form of layers, they should be of great interest for spintronic and other applications.

The calculated total spin magnetic moment is equal to $4\mu_B$ per formula unit and is mainly localized at the Mn atoms. It follows a Slater-Pauling rule similar to the half-metallic Heusler compounds and simple symmetry arguments can explain why there are exactly eleven occupied minority-spin states. The gap originates due to the hybridization of the Te p- and Mn-d states. We expect that the inclusion of Van der Waals corrections, not taken into account in the present study, will give a better description of the interaction between successive layers and will improve the accordance between calculation and experiments in the case of the CsMnTe$_2$ compounds regarding its structural properties. For the other two compounds KMnTe$_2$ and RbMnTe$_2$ it seems that Van der Waals corrections are not crucial for their study.

We believe that the results of our study will increase the interest in these half-metallic ferromagnets since they have also been successfully grown in the past [Kim et al, *Inorganic Chemistry,* vol. 38, pp. 235-242, 1999].

**Acknowledgements**

This work is supported by ***Laboratoire de Recherche : Caractérisation et Valorisation des Ressources Naturelles, Université de Bordj Bou Arreridj, 34000, Algérie***. The authors extend their appreciation to Pr. A. Bentabet for the full support given to complete this work.

Table 1

Calculated Lattice Parameters a and c, c/a ratio, Cell Volumes V, bulk modulus B, first pressure derivative c/a and total energy obtained at the equilibrium volume of XMnTe$_2$ tetragonal structures using GGA compared to experimental data.

|  |  | a (Å) | c (Å) | c/a | V(Å$^3$) | B (GPa) | B' | E (eV) |
|---|---|---|---|---|---|---|---|---|
| **KMnTe$_2$** | NSP | 4.18 | 13.90 | 3.32 | 242.87 | 44 | 3.90 | -417820.550 |
|  | FM | 4.34 | 14.33 | 3.30 | 269.91 | 29 | 4.70 | -417821.828 |
|  | AFM | 4.41 | 14.26 | 3.23 | 277.33 | 28 | 4.05 | -417821.756 |
|  | Exp. [17] | 4.511(4) | 14.909(2) | 3.305 | 303.39(6) | - | - | - |
| **RbMnTe$_2$** | NSP | 4.33 | 13.63 | 3.15 | 255.55 | 41 | 4.42 | -482566.144 |
|  | FM | 4.40 | 14.60 | 3.32 | 282.66 | 27 | 3.83 | -482567.218 |
|  | AFM | 4.44 | 14.63 | 3.30 | 288.41 | 26 | 4.97 | -482567.172 |
|  | Exp. [17] | 4.539(1) | 15.055(2) | 3.317 | 310.17(11) | - | - | - |
| **CsMnTe$_2$** | NSP | 4.60 | 12.36 | 2.69 | 261.54 | 39 | 3.70 | -613419.983 |
|  | FM | 4.76 | 12.91 | 2.71 | 292.51 | 27 | 3.81 | -613421.817 |
|  | AFM | 4.79 | 12.49 | 2.61 | 286.57 | 27 | 4.74 | -613421.431 |
|  | Exp. [17] | 4.5711(4) | 15.527(2) | 3.397 | 324.44(6) | - | - | - |

**Table 2**

Calculated [cal.] and experimental [Exp.] atomic coordinates (*x*, *y*, *z*) for KMnTe$_2$, RbMnTe$_2$ and CsMnTe$_2$ with a space group $I\bar{4}m2$. W.P. (Wyckoff position).

|  | Atoms | W. P. | x Cal. | x Exp. [17] | y Cal. | y Exp. [17] | z Cal. | z Exp. [17] |
|---|---|---|---|---|---|---|---|---|
| **KMnTe$_2$** | K | 2a | 0.0 | 0.0 | 0.0 | 0.0 | 0.0 | 0.0 |
|  | Mn | 2c | 0.0 | 0.0 | 0.5 | 0.5 | 0.25 | 0.25 |
|  | Te | 4e | 0.0 | 0.0 | 0.0 | 0.0 | 0.3574 | 0.3592 |
| **RbMnTe$_2$** | Rb | 2a | 0.0 | 0.0 | 0.0 | 0.0 | 0.0 | 0.0 |
|  | Mn | 2c | 0.0 | 0.0 | 0.5 | 0.5 | 0.25 | 0.25 |
|  | Te | 4e | 0.0 | 0.0 | 0.0 | 0.0 | 0.3511 | 0.3563 |
| **CsMnTe$_2$** | Cs | 2a | 0.0 | 0.0 | 0.0 | 0.0 | 0.0 | 0.0 |
|  | Mn | 2c | 0.0 | 0.0 | 0.5 | 0.5 | 0.25 | 0.25 |
|  | Te | 4e | 0.0 | 0.0 | 0.0 | 0.0 | 0.3429 | 0.3690 |

**Table 3**

Calculated and measured total spin magnetic moments per formula unit ($\mu_{tot}$, in $\mu_B$), calculated spin magnetic moments per atom $\mu_{Mn}$, $\mu_{Te}$ and $\mu_X$ ( X=K, Rb, Cs), the virtual semiconducting gap E$_g$ (in eV), the half-metallic gap E$_{HM}$ (in eV) and the density of states of the majority spin N (up, E$_F$ , in states eV$^{-1}$ spin$^{-1}$ unit-cell$^{-1}$).

|  |  | $\mu_{tot}$ | $\mu_{Mn}$ | $\mu_{Te}$ | $\mu_X$ | $E_g$ | $E_{HM}$ | $N(up, E_F)$ |
|---|---|---|---|---|---|---|---|---|
| KMnTe$_2$ | SP | 4.00 | 3.94 | -0.10 | 0.00 | 1.287 | 0.55 | 1.68 |
|  | Exp. [17] | 4.31 | - | - | - | - | - | - |
| RbMnTe$_2$ | SP | 4.00 | 3.67 | -0.06 | 0.00 | 1.371 | 0.65 | 1.71 |
|  | Exp. [17] | 4.71 | - | - | - | - | - | - |
| CsMnTe$_2$ | SP | 4.00 | 3.62 | -0.04 | 0.00 | 1.253 | 0.57 | 1.54 |
|  | Exp. [17] | 4.86 | - | - | - | - | - | - |

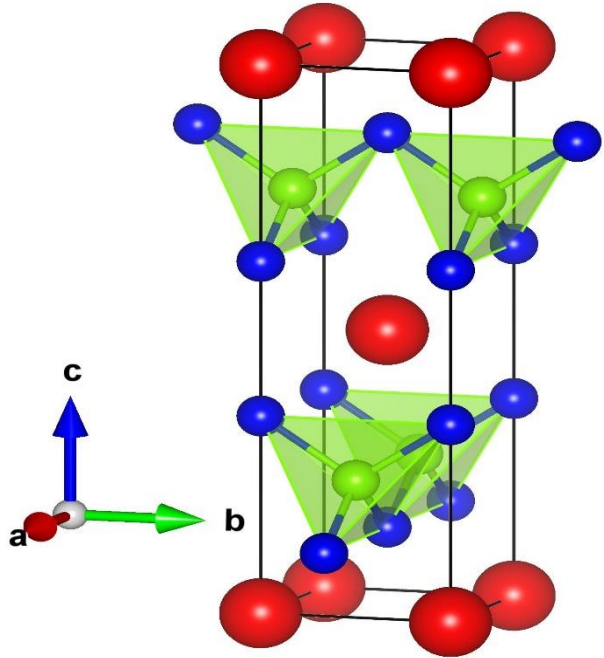

**Fig. 1.** Tetragonal rystal structure of XMnTe$_2$ (X=K, Rb, Cs) layered compounds. The X (alkali atoms) are shown in red sphere, the smal green spheres represent the Mn atoms and the blue sphers are the Te atoms. As shown, the alkali atoms are located between the layers formed by linked MnTe$_4$ tetrahedra.

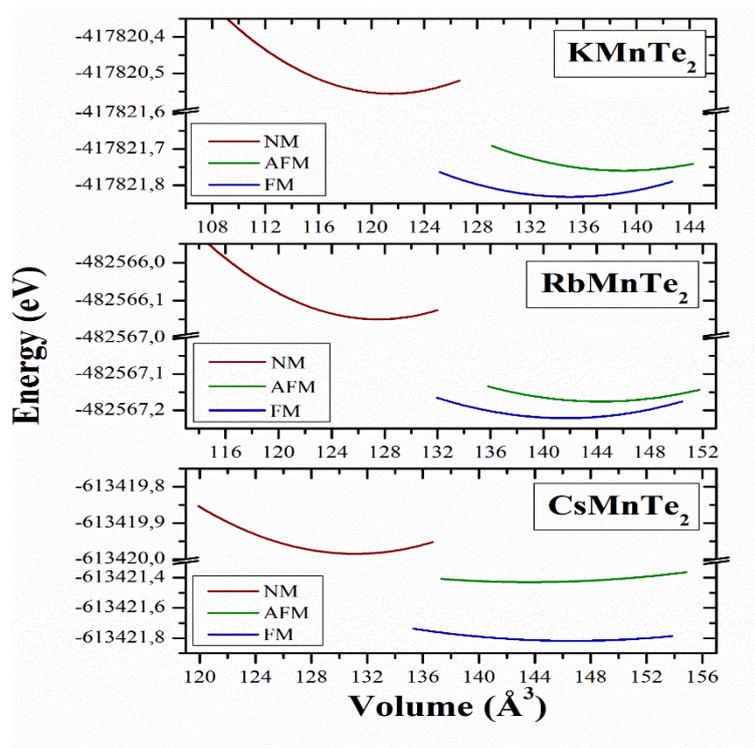

**Fig. 2.** Total energy variation versus volumes for XMnTe2 (X= K, Rb, Cs) structures in ferromagnetic (FM), antiferromagnetic (AFM) and non-magnetic (NM) phases using GGA.

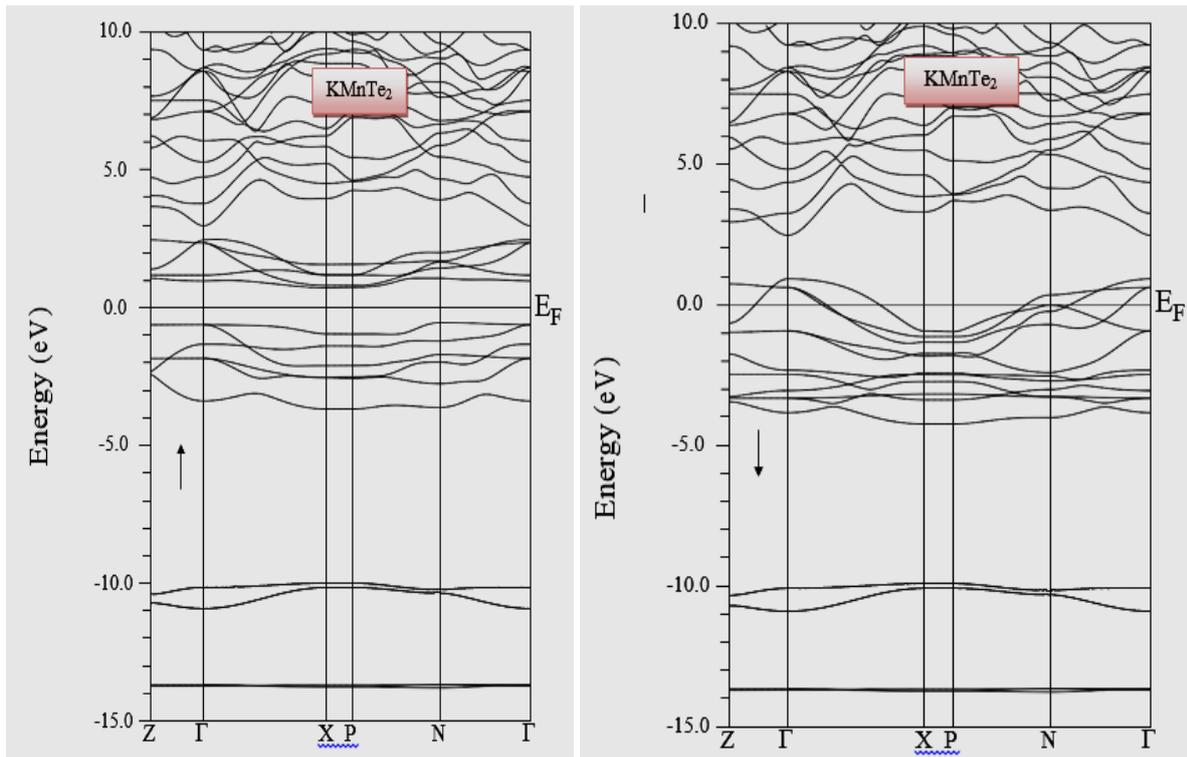

**Fig. 3**. The Spin-polarized band structure of KMnTe$_2$ in ferromagnetic phase (FM): majority spin (↑) and minority spin (↓) using GGA.

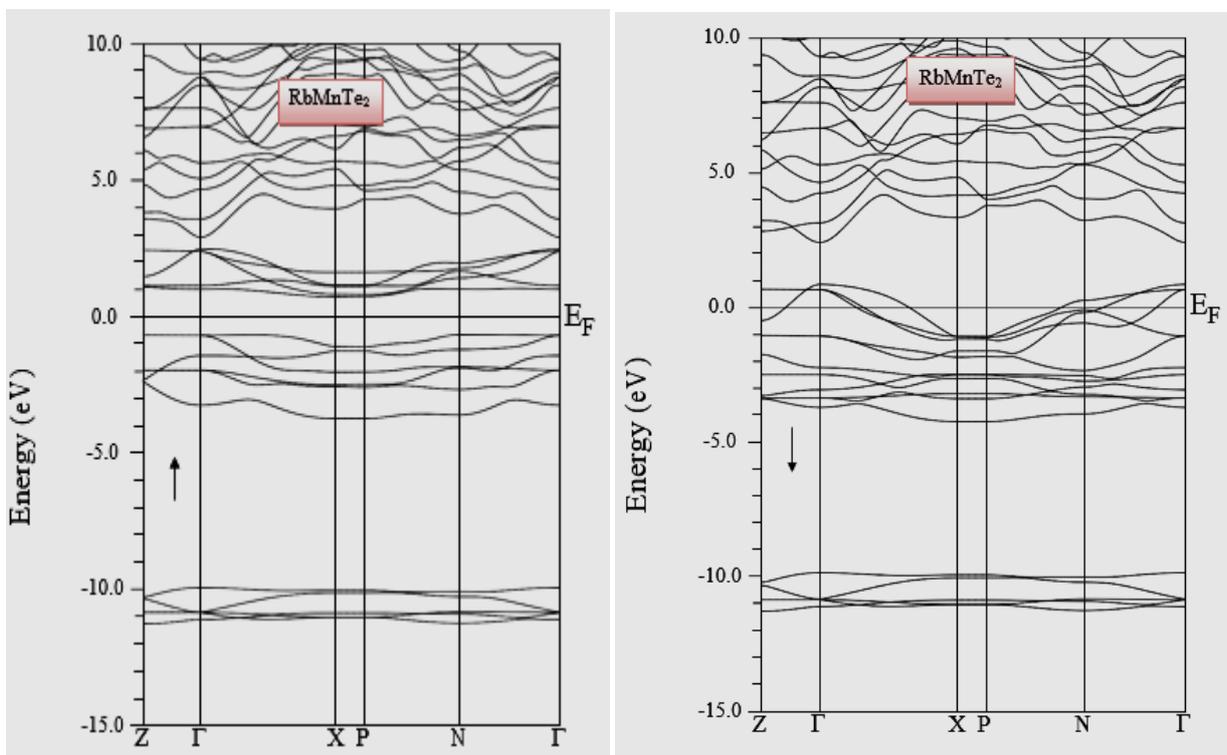

**Fig. 4.** The Spin-polarized band structure of RbMnTe$_2$ in ferromagnetic phase (FM): majority spin (↑) and minority spin (↓).

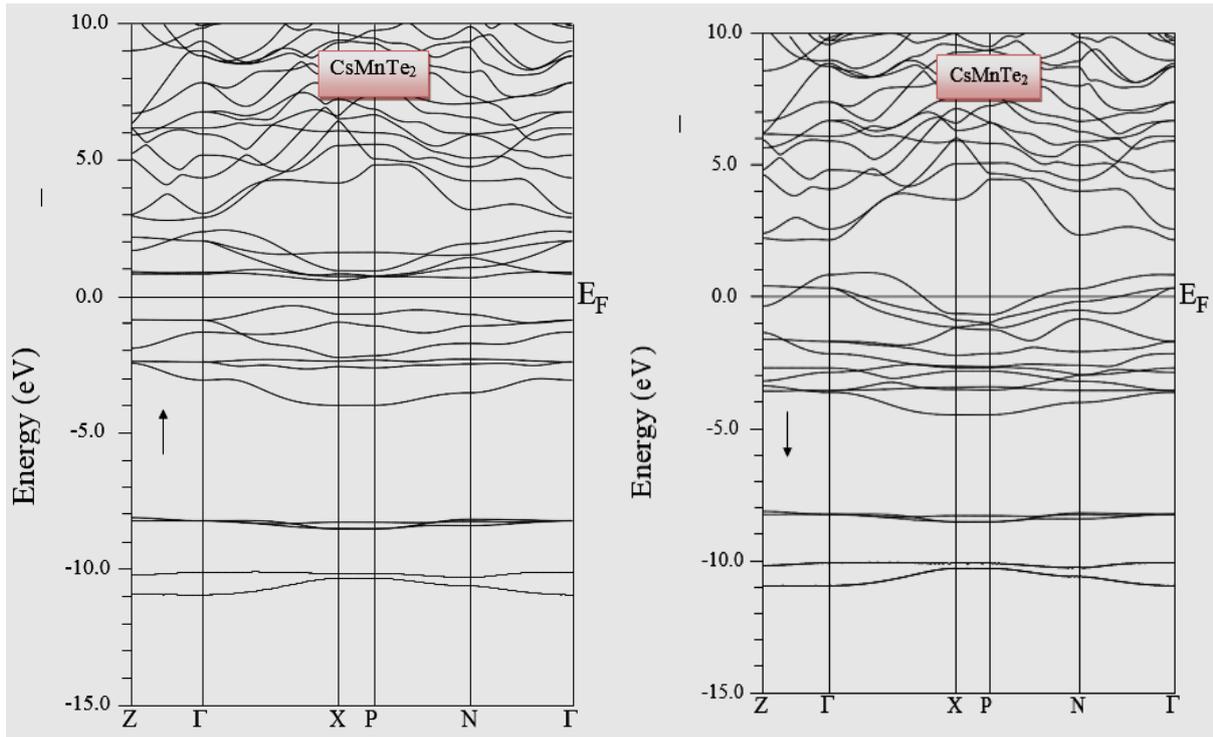

**Fig. 5.** The Spin-polarized band structure of CsMnTe$_2$ ferromagnetic phase (FM): majority spin (↑) and minority spin (↓).

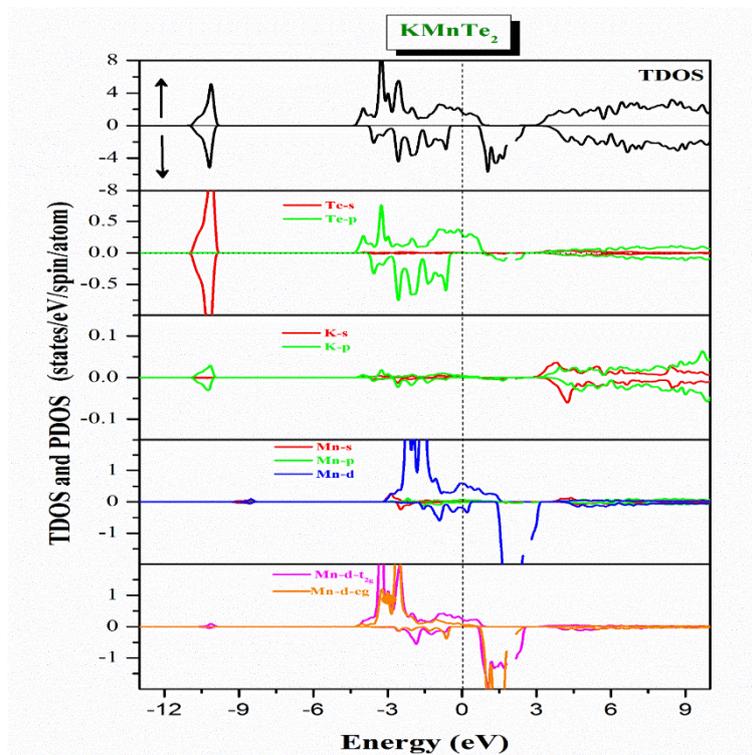

**Fig. 6.** Spin-dependent total and partial density of states for KMnTe$_2$ using GGA. majority spin (↑) and minority spin (↓). The Fermi energy is aligned to zero.

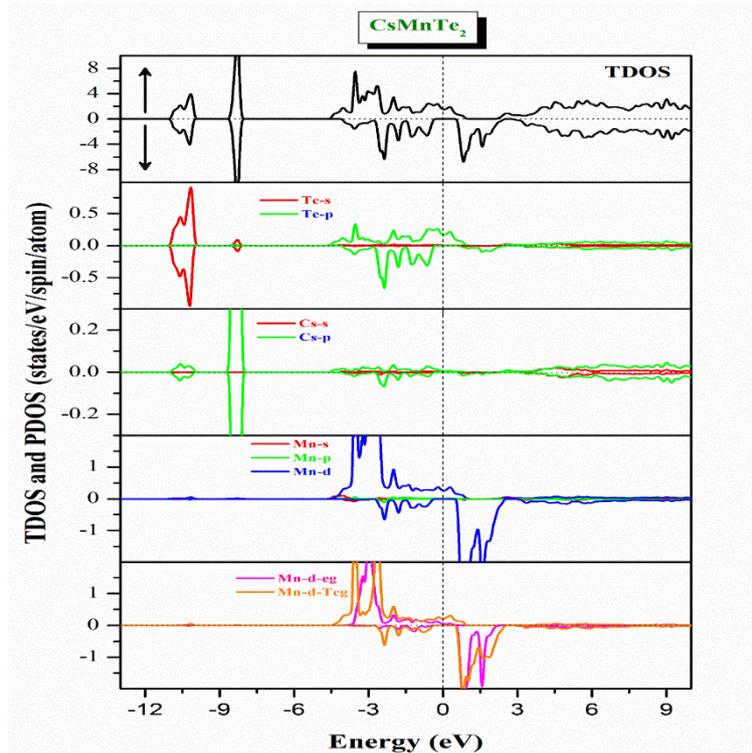

**Fig. 7.** Spin-dependent total and partial density of states for RbMnTe$_2$ using GGA. majority spin (↑) and minority spin (↓). The Fermi energy is aligned to zero.

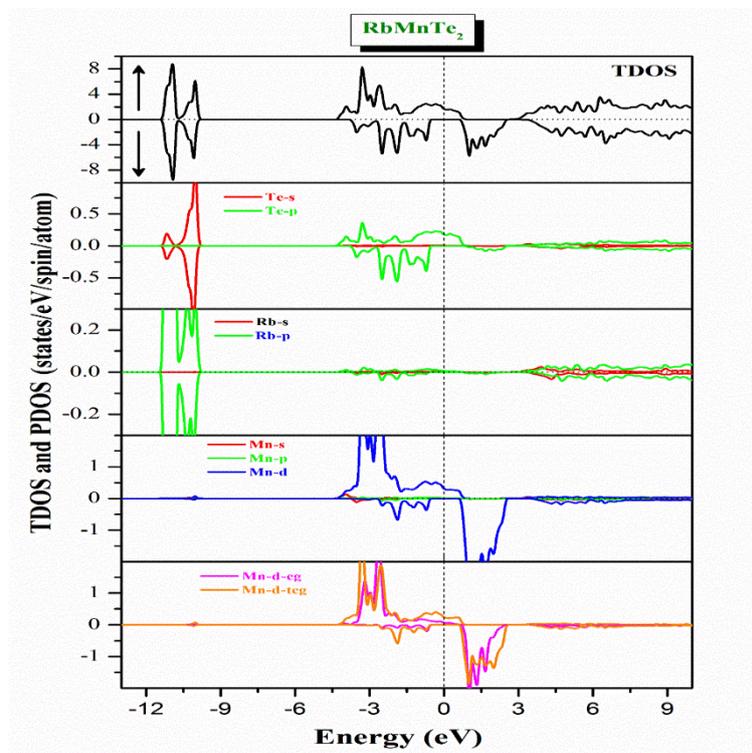

**Fig. 8.** Spin-dependent total and partial density of states for RbMnTe$_2$ using GGA. majority spin (↑) and minority spin (↓). The Fermi energy is aligned to zero.